# Secure mm-Wave Communications with Imperfect Hardware and Uncertain Eavesdropper Location

[1]Saeed Mashdour, Majid Moradikia, Phee Lep Yeoh

**Abstract**

This paper examines the secrecy performance of millimeter-wave (mm-Wave) communications with imperfect hardware and uncertain eavesdropper location. We consider a multiple-antenna source communicating with a single-antenna destination using masked beamforming to transmit the information signals with artificial noise (AN) in the presence of a passive eavesdropper (Eve). For this system, we derive new expressions for the secrecy outage probability (SOP) and secrecy throughput with mm-Wave multipath propagation under slow-fading channel conditions and hardware imperfections. Based on this, optimal power allocation (OPA) solutions are derived for the information and AN signals aimed at minimizing the SOP and maximizing the secrecy throughput. Our results reveal that it is non-trivial to achieve an OPA solution for the general scenario of imperfect hardware. We also highlight that our proposed masked beamforming with OPA scheme significantly enhances the secrecy throughput compared with the benchmark schemes of maximal-ratio transmission and equal power allocation.

## 1. INTRODUCTION

The ubiquity of wireless technology has led to an enormous amount of sensitive and confidential information being transmitted via the open wireless medium. Due to the randomness and broadcast nature of electromagnetic waves, the security of wireless communications conventionally implemented above the physical layer via key based cryptography [1] is becoming increasingly challenging. Recently, the so-called physical layer security (PLS) approach, in which the dynamic characteristics of wireless links are intelligently exploited has attracted significant attention in the literature [2], [3]. For instance, unmanned aerial vehicle (UAV) networks with covert communications for transmission with location uncertainty of the transmitter or receiver was considered in [4] such that the information is not detected by a ground warden. Also, PLS of a UAV network for transmission of confidential information from a base station to multiple receivers utilizing a UAV jammer in presence of multiple eavesdroppers (Eves) was studied in [5]. PLS of overlay device-to-device untrusted relaying in the presence or absence of a friendly jammer was comprehensively analyzed in [6] and secrecy performance of cellular multi-user decode-and-forward based

S. Mashdour and M. Moradikia are with the Department of Electrical and Computer Engineering, Shiraz University, Shiraz, Iran (Email: Smashdour@gmail.com; M.k_majid@yahoo.com); P. L. Yeoh is with the School of Electrical and Information Engineering,The University of Sydney, Sydney, NSW 2006, Australia (email:phee.yeoh@sydney.edu.au)

two-way relaying networks was investigated in [7] where a multi-antenna base station communicates bi-directionally with several users by the aid of decode-and-forward relay(s) under the existence of a multi-antenna passive eavesdropper. In recent years, massive multiple-input multiple-output (MIMO) systems have come into prominence due to the significant performance gains in terms of security, spectral efficiency and energy saving [8], [9]. Typically, multiple antenna transmitters have been considered to transmit narrow directional beams towards the legitimate destination to increase the received signal power relative to the eavesdropper based on known channels state information (CSI) [10]. When the exact CSI of Eve is unknown, the transmitter can broadcast the information signal together with artificial noise (AN) to degrade Eve's channel [11]-[13]. The secrecy performance of AN-aided transmission has been investigated for both fast and slow-fading channels [12], [15]-[17]. For fast-fading channels, the channel coherence time is much shorter than the codeword length, thus beamforming and optimal power allocation (OPA) strategies are based on the ergodic secrecy rate (ESR) [12], [14]. However, in the case of slow-fading channels, outage-based formulations are more suitable. Along this line, the optimal power allocation (OPA) strategy towards the goal of the secrecy outage probability (SOP) minimization is studied in [16], [17]. In recent years, different literatures have used PLS in the context of high-frequency millimeter-wave (mm-Wave) bands, i.e., 30-300 GHz, [18], [19], [21]-[25]. In [21], the ray-cluster channel model has been used which is known to accurately model the multipath propagation in mm-Wave systems. However, only the works of [18], [19], [22] have analyzed the PLS using the ray-cluster channel in mm-Wave communication systems. Another important consideration in wireless systems is the impact of hardware impairments [15], [26], due to I/Q imbalance, oscillator phase noise (PN), high power amplifier (HPA) nonlinearities and quantization errors. It was found in [29], [30] that even after applying transmitter calibration and compensation algorithms at the receiver, there still remains a combination of multiplicative PN and distortion noises at both the transmitter and receiver. Recently, the PLS design with imperfect hardware has been investigated in [15], [31], and [32] for microwave channels. For mm-Wave systems, the effect of imperfect hardware is intensified due to several reasons. Firstly, the high carrier frequency leads to larger multiplication factors which results in phase noise amplification. Also, a bank of phase shifters is needed to connect to high or massive number of antennas to control the hardware cost and power consumption. Furthermore, mm-Wave systems are more sensitive to angular mismatches due to strong directionality provided by large antenna numbers [33]. Thus, it is necessary to consider hardware imperfection in the design and analysis of mm-Wave communication systems.

Millimeter wave systems are known to provide enhanced secrecy against eavesdroppers due to the directional signal beams from highly directional phased-array antennas. However, signal beam reflections from ambient reflectors and eavesdropping strategies using skillfully designed beam exploration have created significant possibilities for eavesdroppers to attack the confidential information [34], so that even only one eavesdropper could intercept transmitted signals successfully [35]. As such, we are motivated in this paper to consider the impact of eavesdropping in the PLS of mm-Wave systems using the ray-cluster model with imperfect hardware and slow-fading channels. To do so, we make the practical assumptions that exact CSI of Eve is unavailable, and the distortion noise follows a Gaussian distribution based on the aggregate contribution of multiple hardware impairments [36]. We assume Eve is randomly located and, based on the slow-fading assumption, there exists a suspicious eavesdropping region where Eve is likely to be located around the source and destination. For this system, we design a new secure AN masked beamforming and separately analyze the OPA aimed at SOP minimization and secrecy throughput maximization. The impact of hardware imperfection results in a more general expression for the mm-wave

OPA compared with the ideal hardware case considered in [18]. It is noted here that due to the larger free-space loss and large number of the implemented antennas in mm-Wave wireless communication systems, perfect knowledge of CSI is difficult to achieve. In such situations, beam training via spatial scanning is a common approach to achieve large beamforming gains by adaptive aligning of the transmitter and the receiver beams. Spatial scanning utilizes exhaustive and hierarchical search strategies through the codebooks that cover the scanning space to determine the best beam which aligns with the dominant paths for communications [37]-[39]. The main contributions of the paper are summarized as follows:

- The generalized system model for transceiver hardware imperfections is detailed after presenting the discrete angular domain channel model for our mm-wave system constructed by spatially resolvable paths.
- We derive the signal-to-noise-plus-distortion ratios (SNDRs) at the Eve and the legitimate destination with secure AN masked beamforming. We also discuss the impact of hardware imperfections in the high-SNR regime.
- We derive a new closed-form expression for the SOP of mm-wave systems with AN and imperfect hardware when the exact eavesdropper location is unknown. Based on this, we formulate the OPA factor between the source and the destination such that the SOP is minimized.
- We also derive the OPA which maximizes the secrecy throughput under a SOP constraint. We also present high SNR analysis to identify the impact of hardware imperfection on the secrecy performance and OPA.
- We show in our analytical and experimental results that imperfect hardware results in more power allocated to the AN signal to boost the secrecy of the system. We also show that the secrecy performance with AN is improved in comparison with traditional maximum ratio transmitting (MRT) beamforming.

Our derived solutions characterize the decrease in secrecy performance of mm-wave systems with imperfect hardware compared to the ideal hardware case. We also highlight that AN beamforming can protect against eavesdropper interception in mm-wave systems with imperfect hardware.

## 2. SIGNAL AND SYSTEM MODEL

In this section, we detail the mm-Wave channel model, transceiver hardware imperfections and dynamic on-off transmission scheme considered in our system model. We assume a multi-antenna source ($S$) equipped with an array of $M$ antennas communicates with a single antenna destination ($D$), while a single antenna Eve ($E$) attempts to eavesdrop the transmitted information. We assume that instantaneous CSI of $D$ is perfectly known at $S$ but the CSI of $E$ is unknown. We consider slow-fading in which the complex channel gains in a single frame are unchanged although they change independently from one frame to another.

### 2.1. Ray-cluster channel model

Due to the sparse scattering and multipath propagation, the mm-Wave channel is well-modeled using a ray-cluster based spatial model, where the channel is represented by multiple clusters each of which comprises several paths. Since it is possible to concentrate the mm-Wave transmission power in a particular cluster, the channel is assumed to have a single cluster. We apply a discrete angular domain channel (DADC) model for the spatially resolvable multipaths [18], [20] where the angular domain at a fixed angular spacing of $\frac{1}{L}$ is uniformly sampled at the transmitter. The length of the mm-Wave transmit antenna array is denoted by $\Delta = \frac{\lambda}{2}$ which determines the angular domain resolvability. Antenna spacing is also denoted by $\Delta = \frac{\lambda}{2}$ where $\lambda$ is the wavelength. Then, the DADC model can be represented by a unitary matrix $\mathbf{W} \in \mathbb{C}^{M \times M}$ whose columns form an orthogonal basis of the transmitted signal space defined as

$$\mathbf{W} \triangleq [\mathbf{R}(\varphi_1), \mathbf{R}(\varphi_2), \ldots, \mathbf{R}(\varphi_M)], \tag{1}$$

where $\varphi_i \triangleq \frac{1}{L}\left(i - 1 - \frac{M-1}{2}\right)$ for $i = 1,2,\ldots,M$ and $\mathbf{R}(\varphi_i) \triangleq \frac{1}{\sqrt{M}}\left[1, e^{-j2\pi\frac{\Delta}{\lambda}\varphi_i}, \ldots, e^{-j(M-1)2\pi\frac{\Delta}{\lambda}\varphi_i}\right]^T$ is the normalized array response (steering vector) at the azimuth angle of departure (AOD) $\theta_i = \sin^{-1}\varphi_i$. Assuming AOD of all paths distributed within the angular range $[\theta_{min}, \theta_{max}]$, where $\{\theta_{min} \leq \theta_{max}\} \in [-\frac{\pi}{2}, \frac{\pi}{2}]$, we will have $\varphi_i \in [\sin\theta_{min}, \sin\theta_{max}]$. Thus, the DADC model is described as

$$\mathbf{h} \triangleq \sqrt{\frac{M\,\alpha}{N}} \mathbf{g}\mathbf{W}^H, \tag{2}$$

where $\mathbf{g} \triangleq [g_1\ g_2\ \ldots\ g_M]$ is the complex channel gain vector. If $\varphi_i \in [\sin\theta_{min}, \sin\theta_{max}]$, $g_i$ is considered as a complex Gaussian coefficient with zero mean and unit variance, otherwise it will be zero. We also denote $\alpha$ as the average path loss between the transmitter and the receiver, and $N$ as number of the spatially resolvable paths ($N < M$). Based on (2), $\mathbf{h}_D \triangleq \sqrt{\frac{M\,\alpha_D}{N_D}}\mathbf{g}_D\mathbf{W}^H$ and $\mathbf{h}_E \triangleq \sqrt{\frac{M\,\alpha_E}{N_E}}\mathbf{g}_E\mathbf{W}^H$ are the $S \rightarrow D$ and $S \rightarrow E$ channels respectively, where $\alpha_D$ and $\alpha_E$ are the corresponding path losses, $\mathbf{g}_D \in \mathbb{C}^{1 \times M}$ and $\mathbf{g}_E \in \mathbb{C}^{1 \times M}$ denote the relevant complex gains, $N_D$ and $N_E$ are numbers of spatially resolvable paths in the destination and the eavesdropper channels with $N_D < M$, $N_E < M$. The basis vectors of the destination and eavesdropper are denoted by $I_{Di}$ for $i = 1,2,\ldots,N_D$ and $I_{Ei}$ for $i = 1,2,\ldots,N_E$, respectively. As shown in Fig. 1, the destination's resolvable paths are selected from the set $\Xi_D \triangleq \{I_{Di} | I_{Di} \in \mathbb{Z}^+, I_{Di} \in [1, M], I_{D1} < I_{D2} < \cdots < I_{DN_D}\}$ while the eavesdropper's resolvable path are selected from the set $\Xi_E \triangleq \{I_{Ei} | I_{Ei} \in \mathbb{Z}^+, I_{Ei} \in [1, M], I_{E1} < I_{E2} < \cdots < I_{EN_E}\}$. We denote the common paths of $D$ and $E$ by the set $\Xi_C \triangleq \Xi_E \cap \Xi_D$. On the other hand, the set $\Xi_A \triangleq \Xi_E \backslash \Xi_C$ shows all the members in $\Xi_E$ but not in $\Xi_C$ and $\Xi_P \triangleq \Xi_D \backslash \Xi_C$ represents all the members in $\Xi_D$ but not in $\Xi_C$. Moreover, For a given matrix $\mathbf{B} \triangleq [\mathbf{b}_1, \mathbf{b}_2, \ldots, \mathbf{b}_m]$ and the set $\Xi \triangleq \{I_i | I_i \in \mathbb{Z}^+, I_i \in [1, m], I_1 < I_2 < \cdots < I_n, n \leq m\}$, the operator $\mathcal{S}(\mathbf{B}, \Xi)$ generates the matrix $\widetilde{\mathbf{B}} \triangleq [\mathbf{b}_{I_1}, \mathbf{b}_{I_2}, \ldots, \mathbf{b}_{I_n}]$ whose columns are selected from $\mathbf{B}$ based on those indices which lie within $\Xi$. Accordingly, we define $\widetilde{\mathbf{g}}_D \triangleq \mathcal{S}(\mathbf{g}_D, \Xi_D) \in \mathbb{C}^{1 \times N_D}$, $\widetilde{\mathbf{g}}_E \triangleq \mathcal{S}(\mathbf{g}_E, \Xi_E) \in \mathbb{C}^{1 \times N_E}$, $\widetilde{\mathbf{W}}_D \triangleq \mathcal{S}(\mathbf{W}, \Xi_D) \in \mathbb{C}^{M \times N_D}$ and $\widetilde{\mathbf{W}}_E \triangleq \mathcal{S}(\mathbf{W}, \Xi_E) \in \mathbb{C}^{M \times N_E}$ through which we can re-express the destination and eavesdropper channels as $\mathbf{h}_D \triangleq \sqrt{\frac{M\,\alpha_D}{N_D}}\widetilde{\mathbf{g}}_D\widetilde{\mathbf{W}}_D^H$ and $\mathbf{h}_E \triangleq \sqrt{\frac{M\,\alpha_E}{N_E}}\widetilde{\mathbf{g}}_E\widetilde{\mathbf{W}}_E^H$, respectively. Additionally, defining $N_{E-C} \triangleq N_E - N_C$ and $N_{D-C} \triangleq N_D - N_C$, the corresponding complex channel gain vectors are defined as $\widehat{\mathbf{g}}_E \triangleq \mathcal{S}(\mathbf{g}_E, \Xi_C) \in \mathbb{C}^{1 \times N_C}$, $\check{\mathbf{g}}_E \triangleq \mathcal{S}(\mathbf{g}_E, \Xi_A) \in \mathbb{C}^{1 \times N_{E-C}}$, $\widehat{\mathbf{g}}_D \triangleq \mathcal{S}(\mathbf{g}_D, \Xi_C) \in \mathbb{C}^{1 \times N_C}$, $\check{\mathbf{g}}_D \triangleq \mathcal{S}(\mathbf{g}_D, \Xi_P) \in \mathbb{C}^{1 \times N_{D-C}}$, as well. It is also noted that based on the intuitive results brought in [18], we can reasonably assume that $N_{D-C} > 0$ and $N_{E-C} > 0$. Since $\mathbf{g}_D$ and $\mathbf{g}_E$ are uncorrelated and independent due to Gaussian distribution, we can also assume that $\widehat{\mathbf{g}}_D$ and $\widehat{\mathbf{g}}_E$ are uncorrelated and independent vectors.

## 2.2. Transceiver hardware imperfection

Fig. 2 shows the statistical behavior of imperfect hardware at node $i \in \{S, D\}$. We consider the ideal eavesdropper case, corresponding to the worst-case condition in terms of secrecy performance, where $E$ does not have any hardware impairments. The distortion noises $\mathbf{\eta}_{tx}$ and $\eta_{rx}$, respectively appeared at the transmitter of $S$ and receiver of $D$ as a consequence of hardware imperfection, causes a mismatch between the desired and the actual transmitted signal. This noise is well-modeled by a Gaussian distributed random variable. We consider the same model as [31] where the variance of the distortion noise at the $i$th node is proportional to the signal power at the corresponding antenna. Accordingly

$$\mathbf{\eta}_{tx} \sim \mathcal{CN}(0, k_{tx}^2 \mathbb{E}\{\mathbf{xx}^H\}) \, , \, \eta_{rx} \sim \mathcal{CN}(0, k_{rx}^2 \mathbb{E}\{|\mathbf{h}_D \mathbf{x}|^2\}) \tag{3}$$

where $\mathbf{x} \in \mathbb{C}^{M \times 1}$ is the transmitted signal vector. The design parameters $k_{tx}$ and $k_{rx}$, so-called error vector magnitudes (EVM), characterize the level of imperfections in the transmitter and receiver hardware, respectively. EVM determines the quality of RF transceivers and is defined as the ratio of the average distortion magnitude to the average signal magnitude. Notably, the EVM measures the joint effect of different hardware imperfections and compensation algorithms and thus it can be measured directly in practice [15], [31]. 3GPP LTE has EVM requirements in the range of $k_{tx}$, $k_{rx} \in [0.08, 0.175]$, where smaller values are needed to achieve higher spectral efficiencies.

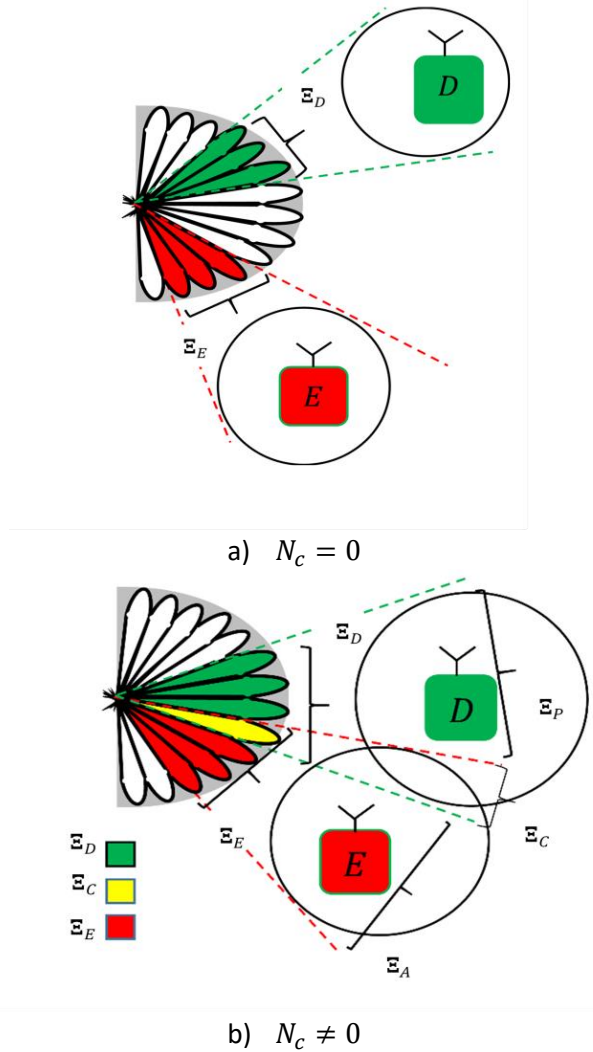

a) $N_c = 0$

b) $N_c \neq 0$

**Figure 1** Model of propagation paths

## 2.3. Dynamic on-off transmission

To address the AN and MRT (as a special case of AN) cases in our slow-fading scenario, we adopt an on-off transmission protocol [17] at the source based on having the instantaneous CSI at the destination and no CSI of the eavesdropper. In the on-off transmission scheme, the transmitter decides to transmit or not based on its knowledge of the receiver's channel so that the transmitter decides when to transmit according to the instantaneous CSI of the destination to prevent undesirable transmissions which incur capacity outages ($R_t > C_D$) or unacceptably high risk of secrecy outage ($C_E > R_e$) where $R_t$ is the codeword rate, $R_e$ is the redundancy rate required to decode the confidential information [40], $Y_D$ and $Y_E$ denote the instantaneous received SINRs, $C_D \triangleq \log_2(1 + Y_D)$ and $C_E \triangleq \log_2(1 + Y_E)$ are the channel capacity of the destination and eavesdropper respectively. According to [16], our on-off transmission scheme is an adaptive scheme where the encoder adaptively selects the transmitted codeword rate $R_t$, according to the instantaneous CSI of the destination's channel. In this scheme, the destination only needs to send its instantaneous SNR $Y_D$ to the transmitter. $R_t$ is arbitrarily set close to $C_D$ which is the largest possible rate without incurring any decoding error at the destination. Let $G \triangleq \|\tilde{\mathbf{g}}_D\|^2$ represents the overall channel gain of the destination, i.e., $G = \hat{G} + \check{G}$ where $\hat{G} \triangleq \|\hat{\mathbf{g}}_D\|^2$, $\check{G} \triangleq \|\check{\mathbf{g}}_D\|^2$ are the instantaneous gains of the common and non-common paths, respectively. Therefore, the source transmits only when $\mathbf{G} \triangleq (\hat{G}, \check{G}) \in \mathbb{R}^2$ is in the transmission region $\xi \triangleq \{\mathbf{G}|C_D > R_s\}$ satisfying the transmission constraint $C_D > R_s$ where $R_s \triangleq R_t - R_e$ is the confidential information rate for a given source codebook.

## 3. SECURE ARTIFICIAL NOISE MASKED BEAMFORMING

We propose a secure artificial noise masked beamforming scheme based on the slow-fading SNDR at the destination. With hardware impairments, it is typically assumed that the distortion noise benefits the system security by degrading the received SNDR at $E$. However, in our ray-cluster channel model, the distortion is an isotropically distributed spatial noise and affects all paths at both $E$ and $D$. As previously mentioned, we assume that while perfect CSI of the destination is known, the exact CSI of the eavesdropper is unavailable at the source. Therefore, in the beamforming transmission design, we assume that partial CSI of the eavesdropper channel should be estimated based on the slow-fading assumption where the exact location of the eavesdropper is unknown but there exists an eavesdropping region around the source and destination. Therefore, we propose to improve the secrecy performance by transmitting artificial interference towards the eavesdropper's resolvable paths which do not include the $N_C$ common paths that overlap with the main channel. As such, a mixed version of information bearing signal and AN is transmitted such that the AN signal is in the null space of the destination's resolvable paths to guarantee not to interfere with the destination's paths.

*Remark 1 (i.i.d. Rayleigh fading channel)*: In typical wireless systems with statistically independent Rayleigh fading channels, if the source has no knowledge of the eavesdropper situation, the AN is broadcasted in the null space of the legitimate channel which is known at $S$. This null space calculation involves the singular value decomposition (SVD) whose complexity is significantly increased by large number of transmit antennas [16]. However, as discussed above, in our case, specific propagation characteristic of mm-Wave channels assist us to transmit AN towards the eavesdropper's resolvable paths excluding the $N_C$ common paths with the destination.■

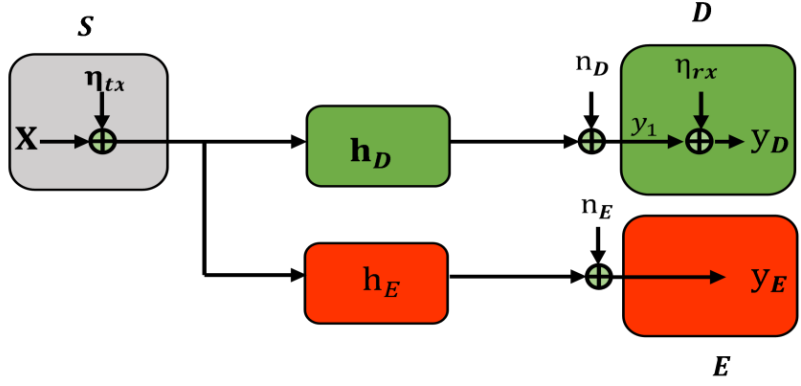

**Figure 2** Block diagram of the system model with imperfections at the source and the destination

Let $s$ represent the unit power information signal, i.e., $\mathbb{E}[|s|^2] = 1$. As such, the signal vector transmitted by $S$ in our proposed scheme is $\mathbf{x} \triangleq \sqrt{\tau P}\mathbf{f}_1 s + \sqrt{\frac{(1-\tau)P}{N_{E-C}}}\mathbf{Fz}$ where $P$ is total transmitted power, $\mathbf{f}_1 \triangleq \mathbf{h}_D^H/\|\mathbf{h}_D\|$ and $\mathbf{F} \triangleq \mathcal{S}(\mathbf{W}, \Xi_A) = [\mathbf{f}_{A_1} \ \mathbf{f}_{A_2} \ \ldots \ \mathbf{f}_{A_{N_{E-C}}}] \in \mathbb{C}^{M \times N_{E-C}}$ with $\mathbb{E}\{\mathbf{F}^H\mathbf{F}\} = \mathbf{I}_{N_{E-C}}$ representing the AN beamforming matrix designed to null out the jamming signal $\mathbb{E}\{\mathbf{zz}^H\} = \mathbf{I}_{N_{E-C}}$ at $D$ i.e., $\mathbf{h}_D\mathbf{F} = 0$. The power allocation factor $\tau \in [0,1]$ determines the power ratio for the information signal and the AN. Interestingly, substituting $\tau = 1$, MRT beamformer is achieved. Using the above definition, the masked beamforming matrix $\mathbf{F}$ is acquired by only choosing specific columns from $\mathbf{W}$ correspoding to $\Xi_A$. According to Fig. 2, the received signals at $D$ and $E$ are

$$y_{D,AN} = \sqrt{\tau P}\mathbf{h}_D\mathbf{f}_1 s + \sqrt{\frac{(1-\tau)P}{N_{E-C}}}\mathbf{h}_D\mathbf{Fz} + \mathbf{h}_D\boldsymbol{\eta}_{tx} + \eta_{rx} + n_D \quad (4)$$

$$y_{E,AN} = \sqrt{\tau P}\mathbf{h}_E\mathbf{f}_1 s + \sqrt{\frac{(1-\tau)P}{N_{E-C}}}\mathbf{h}_E\mathbf{Fz} + \mathbf{h}_E\boldsymbol{\eta}_{tx} + n_E \quad (5)$$

where $n_D$ and $n_E$ are i.i.d. additive white Gaussian noise with zero-mean and variance $\sigma_n^2$, i.e., $n_D \sim \mathcal{CN}(0, \sigma_n^2)$ and $n_E \sim \mathcal{CN}(0, \sigma_n^2)$. According to (4) and (5) after simple algebraic manipulations, the SNDR at $D$ and $E$ can be respectively formulated as

$$\Upsilon_{D,AN} = \frac{\tau d}{\tau e + 1}, \quad (6)$$

$$\Upsilon_{E,AN} = \frac{\tau a u}{(1-\tau)bv + \tau c u + 1}, \quad (7)$$

where $a \triangleq \frac{\beta_E \hat{G}}{G}$, $b \triangleq \frac{\beta_E(1+k_{tx}^2)}{N_{E-C}}$, $c \triangleq k_{tx}^2 a$, $d \triangleq \beta_D G$, $e \triangleq k_{tot}^2 d$, $\beta_D \triangleq \frac{PM\alpha_D}{N_D\sigma_n^2}$, $\beta_E \triangleq \frac{PM\alpha_E}{N_E\sigma_n^2}$, the aggregate level of impairments of all network is denoted by $k_{tot}^2 \triangleq k_{tx}^2 + k_{rx}^2$ and the channel gains are $\sim \text{Gamma}(N_D, 1)$, $\hat{G} \sim \text{Gamma}(N_C, 1)$ and $\check{G} \sim \text{Gamma}(N_{D-C}, 1)$. Furthermore, defining $u \triangleq \left|\hat{\mathbf{g}}_E \frac{\hat{\mathbf{g}}_D^H}{\|\hat{\mathbf{g}}_D\|}\right|^2$, since $\hat{\mathbf{g}}_E$ and the unit norm vector $\frac{\hat{\mathbf{g}}_D^H}{\|\hat{\mathbf{g}}_D\|}$ are both Gaussian and independent, we have $u \sim \exp(1)$ as stated in [20]. We also define $v \triangleq \|\mathbf{g}_E\mathbf{W}^H\mathbf{F}\|^2$ with $v \sim \text{Gamma}(N_{E-C}, 1)$ and the term $\mathbf{g}_E\mathbf{W}^H\mathbf{F}$ therein is equivalent to $\check{\mathbf{g}}_E$. It should also be remarked here that, $u$ and $v$ are independent because $u \in \Xi_C$ and $v \in \Xi_A$ and $\Xi_A \cap \Xi_C = \emptyset$.

*Remark 2: (High-SNR Regime):* To gain deeper insights, we consider (6) and (7) in the high transmit SNR regime. When the transmit power $P$ is sufficiently large, we find that $\Upsilon_{D,AN}$ approaches a constant ceiling

$\Upsilon_{D,AN}^{\infty} = \frac{1}{k_{tot}^2}$ which only depends on the EVMs at the nodes, while the information leaked towards $E$ significantly depends on the power allocation factor $\tau$. This reveals that choosing $\tau$ is crucial for guaranteeing secrecy in our considered network. In addition, it is not possible to achieve some target secrecy rates due to the ceiling defined above, no matter how much large the value of transmit power $P$. ∎

*Remark 4: (Ideal Hardware-Masked Beaforming):* For the special assumption of ideal hardware at the nodes, the SNDRs calculated in (6) and (7) can be simplified to the received SNRs in [18], [22] given by

$$\Upsilon_{D,AN}^{id} \triangleq \tau d, \tag{8}$$

$$\Upsilon_{E,AN}^{id} \triangleq \frac{N_{E-C} \tau a u}{(1-\tau)\beta_E v + N_{E-C}}, \tag{9}$$

Comparing (8), (9) with (6), (7), it can be clearly seen that the terms $\tau e$ and $\tau cu$ in (6) and (7), respectively, contribute to a more complicated expression for the non-ideal hardware SNDRs. Based on (6), (7), we see that the distortion noises emitted by $S$ results in interference at both $E$ and $D$, while the self-distortion noise at $D$ only degrades the received SNDR at the legitimate destination. Thus, even though hardware imperfection can intrinsically confuse passive eavesdroppers, the secrecy performance is still degraded. ∎

it is also worthy of mention that the eavesdropper's SNDR in both cases of ideal and non-ideal hardware, which were derived respectively in (7) and (9), is calculated based on the fact that $\mathbf{g}_E \mathbf{g}_D^H = \hat{\mathbf{g}}_E \hat{\mathbf{g}}_D^H$ which is only achieved when the $i$th path is chosen from the common set $\Xi_C$, i.e., $i \in \Xi_C$, $\mathbf{g}_{Ei} \mathbf{g}_{Di}^* \neq 0$. If there are no common paths between $D$ and $E$, i.e., $N_C = 0$, the array steering vectors within $\widetilde{\mathbf{W}}_D$ and $\widetilde{\mathbf{W}}_E$ will be orthogonal, resulting in $\Upsilon_{E,AN} = 0$. Thus, it can be concluded that in the situation of $N_C = 0$, the secrecy can be effectively guaranteed by using a simple beamformer which has low implementation complexity in contrast to other complicated precoder techniques at the multi-antenna transmitter.

It is also worthy of mention that if $N_C$ is increased, $\hat{G}$ in both the numerator and denominator of $\Upsilon_{E,AN}$ will increase. Thus, increasing the number of common paths between $E$ and $D$ results in more information likely to be leaked towards $E$ and a higher amount of distortion noise due to imperfection injected into $N_C$. Therefore, unlike the perfect hardware case where the intercept probability is optimistically assumed to be only affected by the relationship between spatially resolvable paths of $E$ and $D$, we highlight that the level of hardware imperfections can significantly influence the secrecy performance.

The secrecy performance is evaluated from two perspectives. In section 4, the primary design objective is to minimize the SOP. To do so, a closed form expression for the SOP is obtained and the optimal power allocation factor $\tau_{opt}^*$ is derived through minimizing the SOP. In the second approach presented in Section 5, the OPA that maximizes the throughput $\tau_{to}^*$ subject to a maximum permissible SOP constraint is derived.

## 4. SOP ANALYSIS AND THE CORRESPONDING OPA STRATEGY

With the on-off transmission strategy described in Section 2, the SOP is expressed as $\mathcal{P}_{SO} \triangleq \mathbb{P}_r\{C_D - C_E < R_s | \mathbf{G} \in \xi\}$ for a fixed rate $R_s$ based on a common codebook at $S$ and $D$. Utilizing on-off transmission, the source stops transmissions unless the instantaneous destination's channel gain $G$ exceeds the threshold $\mu$ which is determined relying on the conditions of the transmission region $\xi$. Thus, to guarantee a positive $R_e$ against the eavesdropper the minimum value for the power allocation factor is determined according to the transmission region described in dynamic on-off transmission, where, $C_D = \log_2(1 + Y_{D,AN}) > R_s$. Therefore, using (6) we can conclude that

$$Y_{D,AN} = \frac{\tau d}{\tau e + 1} > (T - 1) \Rightarrow \tau > \tau_{min} \triangleq \frac{\bar{T}}{d - e\bar{T}}, \tag{10}$$

where $\bar{T} \triangleq T - 1$ with $T \triangleq 2^{R_s}$ and therefore the overall channel gain threshold is $G > \mu = \frac{(T-1)(\tau e+1)N_D \sigma_n^2}{PM\alpha_D}$. On the other hand, based on $Y_{D,AN}$ in (6), the SOP is given by

$$\mathcal{P}_{SO} = \mathbb{P}_r\left\{Y_{E,AN} > \frac{Y_{D,AN} - \bar{T}}{T}\right\} = \mathbb{P}_r\left\{Y_{E,AN} > \frac{\tau d - \bar{T}(\tau e + 1)}{T(\tau e + 1)}\right\}, \tag{11}$$

In order to examine (11), we derive the cumulative distribution function (CDF) of $Y_{E,AN}$ given by

$$F_{Y_{E,AN}}(x) \triangleq \mathbb{P}_r\left\{\frac{\tau a u}{(1-\tau)bv + \tau c u + 1} < x\right\} = \mathbb{P}_r\left\{u < \frac{x((1-\tau)bv + 1)}{\tau(a - cx)}\right\}, \tag{12}$$

and exploit the following helpful lemma.

*Lemma 1:* The cdf of r.v. $\hat{\Lambda} \triangleq \frac{\alpha_1 \Lambda}{\alpha_2 \Lambda + \alpha_3}$, where $\Lambda$ is a nonnegative r.v. and $\alpha_i|_{i=1}^3$ represent positive constants, is obtained as

$$F_{\hat{\Lambda}}(x) = \begin{cases} F_\Lambda\left(\frac{\alpha_3 x}{\alpha_1 - \alpha_2 x}\right) & ; 0 \leq x < \frac{\alpha_1}{\alpha_2} \\ 1 & ; \quad x \geq \frac{\alpha_1}{\alpha_2} \end{cases}, \tag{13}$$

*Proof*: From the definition of CDF we have

$$F_{\hat{\Lambda}}(x) = \mathbb{P}_r\left\{\frac{\alpha_1 \Lambda}{\alpha_2 \Lambda + \alpha_3} \leq x\right\} = \mathbb{P}_r\{\Lambda(\alpha_1 - \alpha_2 x) \leq \alpha_3 x\}, \tag{14}$$

If $(\alpha_1 - \alpha_2 x) < 0$, the probability of (14) becomes one, while in the case of having $(\alpha_1 - \alpha_2 x) \geq 0$, it will be equal to $F_\Lambda\left(\frac{\alpha_3 x}{\alpha_1 - \alpha_2 x}\right)$.

Using lemma 1, $F_{Y_{E,AN}}(x)$ is derived as

$$F_{Y_{E,AN}}(x) = \begin{cases} F_u\left(\frac{x((1-\tau)bv+1)}{\tau(a-cx)}\right) & ; 0 \leq x < \frac{1}{k_{tx}^2} \\ 1 & ; \quad x \geq \frac{1}{k_{tx}^2} \end{cases}, \tag{15}$$

where $F_u\left(\frac{x((1-\tau)bv+1)}{\tau(a-cx)}\right)$ is acquired as

$$F_u\left(\frac{x((1-\tau)bv+1)}{\tau(a-cx)}\right) = 1 - \mathbb{E}_v\left[e^{-\frac{(1-\tau)bx}{\tau(a-cx)}v - \frac{x}{\tau(a-cx)}}\right] = 1 - e^{-\frac{x}{\tau(a-cx)}} \int_0^\infty \frac{v^{N_{E-C}-1}}{\Gamma(N_{E-C})} e^{-\left(1 + \frac{(1-\tau)bx}{\tau(a-cx)}\right)v} dv$$

$$= 1 - e^{-\frac{x}{\tau(a-cx)}}\left[1 + \frac{(1-\tau)bx}{\tau(a-cx)}\right]^{-N_{E-C}}, \tag{16}$$

Now, given (11) the SOP is given by

$$\mathcal{P}_{SO} = \begin{cases} \mathcal{P}_{SO}^{\circ}(\tau) & ; \gamma_1(\tau) < T < \gamma_2(\tau) \\ 0 & ; T \leq \gamma_1(\tau) \end{cases}, \quad (17)$$

where $\gamma_1(\tau) \triangleq \frac{\tau d k_1 + k_{tx}^2}{\tau d k_2 + k_{tx}^2 + 1}$ and $\gamma_2(\tau) \triangleq \frac{\tau d k_3 + 1}{\tau d k_{tot}^2 + 1}$ with $k_1 \triangleq k_{tx}^2 + k_{tx}^2 k_{tot}^2$, $k_2 \triangleq k_{tot}^2 + k_{tx}^2 k_{tot}^2$, $k_3 \triangleq k_{tot}^2 + 1$ and $\mathcal{P}_{SO}^{\circ}$ is given by

$$\mathcal{P}_{SO}^{\circ}(\tau) \triangleq \exp\left(-\frac{\tau d - (\tau e + 1)\bar{T}}{\tau\big((\tau e + 1)Ta - c(\tau d - (\tau e + 1)\bar{T})\big)}\right)$$

$$\times \left[1 + \frac{(1-\tau)b(\tau d - (\tau e + 1)\bar{T})}{\tau\big((\tau e + 1)Ta - c(\tau d - (\tau e + 1)\bar{T})\big)}\right]^{-N_E - C}, \quad (18)$$

Our new expression in (18) generalizes the SOP expression derived for the ideal hardware case in [18]. It should be remarked that, since we use on-off transmission scheme with regard to the fact that $0 \leq \tau_{min} \leq 1$ we have to consider two further cases for the SOP. Therefore, the cases of $T > \gamma_3$ with $\gamma_3 \triangleq \frac{k_{tot}^2 + 1}{k_{tot}^2}$ and $< \frac{\bar{T}}{\beta_D(1 - k_{tot}^2 \bar{T})}$, should be taken into account where $S$ stops transmitting if one of these conditions holds true. From the practical point of view, the condition $T > \gamma_3$ implies that unlike perfect hardware situation where the design parameter $R_s$ is independent of transceivers hardware qualities, in our generalized considered case, $R_s$ should be determined in accordance with the aggregate level of impairments imposed on the whole network, before the data transmission. Then, even if $T < \gamma_3$, the source node examines the overall destination's channel feedback to determine if the constraint $G < \frac{\bar{T}}{\beta_D(1 - k_{tot}^2 \bar{T})}$ (or equivalently $T > \frac{dk_3 + 1}{1 + dk_{tot}^2} = \gamma_2(1)$) is satisfied and full power is assigned for AN transmission (i.e., $\tau = 0$) or the source totally suspends its transmission.

Based on what was discussed above together with (17), finally the overall SOP is derived as

$$\mathcal{P}_{SO}^* = \begin{cases} 1 & T > \gamma_3 \\ \mathcal{P}_{SO}^{\circ}(\tau_{opt}^*) & T < \gamma_3 \,\&\, T < \gamma_2(1) \,\&\, \gamma_1(\tau_{opt}^*) < T < \gamma_2(\tau_{opt}^*), \\ 0 & T < \gamma_3 \,\&\, T < \gamma_2(1) \,\&\, T < \gamma_1(\tau_{opt}^*) \end{cases} \quad (19)$$

where the SOP is 1 for target secrecy rates above a threshold determined by the imperfection of the nodes, i.e., $> \gamma_3$. When $T < \gamma_3$, the SOP approaches zero with increasing $P$. Our result is different from the ideal hardware scenario in which the SOP, regardless to any target secrecy rate, always approaches zero with increasing SNR [18], [41].

Based on (19), $\mathcal{P}_{SO}(\tau^*)$ is attained by substituting the OPA factor $\tau_{opt}^*$, which minimizes the SOP in (17). From (19) it can be inferred that if the predefined target secrecy level satisfies $T < \gamma_3$, the OPA that minimizes the SOP can be derived. Minimizing the instantaneous SOP, with respect to the power allocation factor $\tau$ yields the optimum value $\tau_{opt}^*$ as follows

$$\begin{aligned} \tau_{opt}^* &= \arg\min_\tau \mathcal{P}_{SO}(\tau) \\ \text{s.t. } &\tau_{min} < \tau < 1 \end{aligned} \quad (20)$$

The minimization of $\mathcal{P}_{SO}(\tau)$ is equivalent to minimizing $\mathcal{P}_{SO}^{\circ}(\tau)$. However, because of the complicated mathematical structure of $\mathcal{P}_{SO}^{\circ}(\tau)$, it is non-trivial to simply investigate the monotonicity of $\mathcal{P}_{SO}(\tau)$. As an alternative solution, since we have $\mathcal{P}_{SO}(\tau) \triangleq \mathbb{P}_r\left\{\log_2\left(\frac{1 + Y_D(\tau)}{1 + Y_E(\tau)}\right) < R_s \Big| G \in \xi\right\}$ and also given that $\log(\cdot)$ is a

monotonically increasing function, the minimization of $\mathcal{P}_{SO}$ is equivalent to the maximization of $\phi(\tau) \triangleq \frac{1+\Upsilon_D(\tau)}{1+\Upsilon_E(\tau)}$ as follows

$$\tau_{opt}^* = \arg\max_\tau \phi(\tau) \qquad\qquad (21)$$
$$\text{s.t. } \tau_{min} < \tau < 1$$

Substituting $\Upsilon_{D,AN}$ and $\Upsilon_{E,AN}$, which was earlier obtained respectively in (6) and (7), $\phi(\tau)$ can be reformulated as

$$\phi(\tau) \triangleq \frac{c_1\tau^2 + c_2\tau + c_3}{c_4\tau^2 + c_5\tau + c_6} \qquad\qquad (22)$$

where $c_1 \triangleq (e+d)(cu - bv)$, $c_2 \triangleq (e+d)(bv+1) + cu - bv$, $c_3 \triangleq bv + 1 = c_6$, $c_4 \triangleq e(cu + au - bv)$ and $c_5 \triangleq e(bv+1) + cu + au - bv$. To obtain optimal points of $\phi(\tau)$ over $\tau$, we first investigate the convexity of $\phi(\tau)$. To do so, the first derivative of $\phi(\tau)$ with respect to the $\tau$ is given by

$$\frac{\partial \phi(\tau)}{\partial \tau} = \frac{\varepsilon_1\tau^2 + \varepsilon_2\tau + \varepsilon_3}{(c_4\tau^2 + c_5\tau + c_6)^2}, \qquad\qquad (23)$$

where $\varepsilon_1 \triangleq c_1c_5 - c_2c_4$, $\varepsilon_2 \triangleq 2c_3(c_1 - c_4)$ and $\varepsilon_3 \triangleq c_3(c_2 - c_5)$. Since the denominator is always positive, we can readily find that the sign of $\frac{\partial \phi(\tau)}{\partial \tau}$ follows that of the numerator $\Omega(\tau) \triangleq \varepsilon_1\tau^2 + \varepsilon_2\tau + \varepsilon_3$. In other words, examining the sign of $\Omega(\tau)$ is sufficient to investigate the monotonicity of $\phi(\tau)$. Now, relying on corollary 1 in [15] together with the sign of $\Omega(\tau)$ at the boundaries $\tau = \tau_{min}$ and $\tau = 1$, the following proposition is presented to solve the maximization of $\phi(\tau)$ in (21).

*Proposition 1: Based on the sign of $\Omega(\tau)$ at the boundaries $\tau = \tau_{min}$ and $\tau = 1$, $\phi(\tau)$ may be convex, concave or it is possible to be neither convex nor concave in the feasible set $\tau \in (\tau_{min}, 1]$. Following this fact, the optimal point is given by*

$$\tau_{opt}^* = \begin{cases} \mathcal{T}\{\max\{\Omega(\tau_{min}), \Omega(\tau_1^\circ), \Omega(\tau_2^\circ), \Omega(1)\}\} & ; \Omega(\tau_{min}).\Omega(1) > 0 \\ \mathcal{T}\{\max\{\Omega(\tau_{min}), \Omega(1)\}\} & ; \Omega(\tau_{min}) < 0 \,\&\, \Omega(1) > 0, \\ \tau_1^\circ & ; \Omega(\tau_{min}) > 0 \,\&\, \Omega(1) < 0 \end{cases} \qquad (24)$$

where $\tau_1^\circ = -\left(\varepsilon_2 + \sqrt{\varepsilon_2^2 - 4\varepsilon_1\varepsilon_3}\right)/(2\varepsilon_1)$ and $\tau_2^\circ = -\left(\varepsilon_2 - \sqrt{\varepsilon_2^2 - 4\varepsilon_1\varepsilon_3}\right)/(2\varepsilon_1)$ are the zero-crossing points of $\Omega(\tau)$. The operator $\mathcal{T}\{f(x_i)\}$ is defined as a function which extracts the point $x_i$ corresponding to $y_i \triangleq f(x_i)$. Please refer to Appendix A for the proof.

## 5. SECRECY THROUGHPUT ANALYSIS AND CORRESPONDING OPA STRATEGY

In the previous section, we assume that the code rates are fixed and do not vary with **G**. In this section, a different approach is considered where the power allocation factor $\tau$ is optimized for different code rates, i.e., the transmission code rate $R_t$ and source code rate $R_s$ are adapted for each channel realization of $D$. This leads to a higher throughput compared to the previous scheme at the expense of higher complexity. Hence, in this second approach the code rates $R_t$ and $R_s$ are potentially functions of **G**, i.e., $R_t(\mathbf{G})$ and $R_s(\mathbf{G})$. Based on on-off transmission scheme, when $G \in \mathbf{G}$, the average secrecy throughput is defined as $\Upsilon \triangleq \mathbb{E}_G\{R_s(\mathbf{G})\}$ while in the case of $G \notin \mathbf{G}$ we have $R_s(\mathbf{G}) = 0$ [12]. Our aim is to maximize the throughput under a maximum tolerable SOP constraint. To do so, the maximization of $\Upsilon$, which is equivalent to the maximization of $R_s(\mathbf{G})$ for each destination's channel realization, is accomplished over the parameters $R_t$ and power allocation factor $\tau$ to satisfy the requirement on the SOP, i.e., $\mathcal{P}_{SO}(\mathbf{G}) < \varepsilon$. The throughput optimization problem is formulated as

$$\begin{aligned} \tau_{to}^* &= \arg\max_{\tau, R_t} R_s(\mathbf{G}) \\ \text{s.t. } &0 < R_s(\mathbf{G}) < R_t(\mathbf{G}) \leq C_D \\ &\mathcal{P}_{SO}(\mathbf{G}) \leq \varepsilon \\ &0 \leq \tau \leq 1 \end{aligned} \quad (25)$$

where the first constraint guarantees to have a reliable link to the intended receiver i.e., corresponding transmission region of the second OPA strategy. The second condition imposes secrecy constraint in which the parameter $0 \leq \varepsilon \leq 1$ is a predefined threshold that indicates the maximum tolerable SOP. The third condition is the power allocation constraint. By relaxing the optimization problem (25), it can be rewritten as follows, whereas the derivation is relegated to Appendix B.

$$\begin{aligned} \tau_{to}^* &= \arg\max_\tau R_s(\tau) \\ \text{s.t. } &\frac{d}{\tau e + 1} \geq k(\tau), \ 0 < \tau < 1 \end{aligned} \quad (26)$$

where $k(\tau) \triangleq \frac{F_{Y_{E,AN}}^{-1}(1-\varepsilon)}{\tau}$ and $R_s(\tau)$ is given by

$$R_s(\tau) \triangleq \log_2\left(\frac{1 + Y_{D,AN}}{1 + \tau k(\tau)}\right) = \log_2\left(\frac{\tau(d+e) + 1}{\tau^2 e k(\tau) + \tau(k(\tau) + e) + 1}\right), \quad (27)$$

Notably, the transmission region constraint $\xi_{to} = \{\mathbf{G}|\frac{d}{\tau e + 1} \geq k(\tau)\}$ in (26) implies that $S$ only transmits when this condition is satisfied.

To solve the maximization problem in (26), we have to first investigate the behavior of $k(\tau)$. Using (16) and the definition of $k(\tau)$, the equation $F_{Y_{E,AN}}(\tau k(\tau)) = 1 - \varepsilon$ can be reformulated as follows

$$Q(k) \triangleq \exp\left(-\frac{k(\tau)}{a - c\tau k(\tau)}\right)\left[1 + \frac{(1-\tau)k(\tau)}{a - c\tau k(\tau)}\right]^{-N_E - C} - \varepsilon \quad (28)$$

where $k(\tau)$ is a monotonically increasing function in the feasible set $0 \leq \tau \leq 1$, (See Appendix C for the proof). From (28) the maximum value of $k(\tau)$, denoted by $k_{max}$, is attained when $\tau = 1$ as

$$k_{max} = k(1) = -a\ln(\varepsilon)/(1 + c\ln(\varepsilon)) \quad (29)$$

where $Q(k)$ is a monotonically decreasing function of $k$ (please refer to Appendix D for the proof). Thus, $Q(k = 0) = Q_{max} = 1 - \varepsilon \geq 0$. Also, $Q_{min}$ is acquired as

$$Q_{min} = Q(k_{max}) = e^{\frac{\ln(\varepsilon)}{1 + (1+\tau)\ln(\varepsilon)}}\left[1 - (1 - \tau)\frac{\ln(\varepsilon)}{1 + (1+\tau)\ln(\varepsilon)}\right]^{-(N_E - c)} - \varepsilon \quad (30)$$

In the above equation, $e^{\frac{\ln(\varepsilon)}{1+(1+\tau)\ln(\varepsilon)}} \leq e^{\ln(\varepsilon)} = \varepsilon$. Thus, $Q_{min} \leq \varepsilon\left[1 - (1-\tau)\frac{\ln(\varepsilon)}{1+(1+\tau)\ln(\varepsilon)}\right]^{-(N_E - N_C)} - \varepsilon \leq 0$. We can easily conclude that $Q(k)$ has a unique zero crossing point. Applying the bisection method we can find a value $k$ which is the zero crossing point of $Q(k)$ within the range $[0, k_{max}]$ for a given $\tau$. To start bisection method, upper and lower bound of $k$ are considered as $[k_l, k_u] = [0, k_{max}]$. To solve (26), optimal value of $\tau$ should be found such that $R_s$ is maximized. The optimal value of $\tau$ is derived for two cases of $R_s(1)$.

*Case 1)* $R_s(\tau)$ *is a concave function of* $\tau$. In this case, the optimal value is given by

$$\tau_{to}^* = \begin{cases} 1, & Z > 0 \\ \tau^*, & otherwise \end{cases} \tag{31}$$

where $\tau^*$ is the unique root of the following equation

$$\frac{dR_s}{d\tau} = \frac{1}{ln2}\left(\frac{1}{(\tau e + 1)(\tau(e+d)+1)} - \frac{k + \tau\frac{\partial k}{\partial \tau}}{1+\tau k}\right) = 0 \tag{32}$$

and $Z = \frac{1}{ln2}\left(\frac{d}{(e+1)(e+d+1)} + \frac{(N_{E-C}+1)\, a\, ln(\varepsilon)\left(1+\frac{c\, ln(\varepsilon)}{1+c\, ln(\varepsilon)}\right)}{1+(c-a)ln(\varepsilon)}\right)$.

*Case 2)* $R_s(\tau)$ *is a non-concave function of* $\tau$. In this case, the optimal value is given by

$$\tau_{to}^* = \begin{cases} \begin{cases} \tau_1 & R_s(\tau_1) > R_s(1) \\ 1 & R_s(\tau_1) < R_s(1) \end{cases}, & Z > 0 \\ \begin{cases} \tau_1' & R_s(\tau_1') > R_s(\tau_3) \\ \tau_3 & R_s(\tau_1') < R_s(\tau_3) \end{cases}, & otherwise \end{cases} \tag{33}$$

where, $\tau_1$, $\tau_1'$ and $\tau_3$ are discussed in Appendix E. Therefore, $\tau_{to}^*$ could be obtained using bisection method. As such, $S$ only transmits when $\mathbf{G}\epsilon\xi_{to}$. Otherwise, $S$ keeps silent to guarantee a positive $R_s$ and the only power allocation factor which could meet the transmission condition is the optimal power allocation that maximizes $R_s(\mathbf{G})$. Therefore, the maximum secrecy throughput is defined as

$$\Upsilon = \iint_{\xi_{to}} R_{sopt}(\mathbf{G}) f_{\breve{G},\hat{G}}(x,y)\, dxdy \tag{34}$$

where $R_{sopt}(\mathbf{G})$ is the maximized secrecy rate for $\tau_{to}^*$ and $f_{\breve{G},\hat{G}}(x,y)$ is the joint probability density function (pdf) of $\breve{G}$ and $\hat{G}$. Since $\Xi_P$ and $\Xi_C$ have no common path, $f_{\breve{G},\hat{G}}(x,y)$ could be replaced by $f_{\breve{G}}(x) \times f_{\hat{G}}(y)$. Similar to what is mentioned for $v$ in section 3.2, $\hat{G} \sim Gamma(N_C, 1)$ and $\breve{G} \sim Gamma(N_{D-C}, 1)$. Therefore, (34) can be rewritten as

$$\Upsilon = \iint_{\xi_{to}} R_{sopt}(\mathbf{G}) f_{\breve{G}}(x) f_{\hat{G}}(y)\, dxdy \tag{35}$$

where $f_{\breve{G}}(x) = \frac{e^{-x} x^{N_{D-C}-1}}{(N_{D-C}-1)!}$ and $f_{\hat{G}}(y) = \frac{e^{-y} y^{N_C-1}}{(N_C-1)!}$. The secrecy throughput for MRT beamforming can be derived as a special case with $\tau = 1$ and $k = k_{max}$ given in (29). Thus, $R_{sopt}(\mathbf{G})$ with MRT beamforming is derived as

$$R_{sopt}(\mathbf{G}) = \log_2\left(\frac{1+\frac{d}{e+1}}{1-\frac{a\ln\varepsilon}{1+c\ln\varepsilon}}\right) \tag{36}$$

Redefining $a = \bar{a}\frac{\hat{G}}{\hat{G}+\check{G}}$, $c = \bar{c}\frac{\hat{G}}{\hat{G}+\check{G}}$, $d = \bar{d}(\hat{G}+\check{G})$ and $e = \bar{e}(\hat{G}+\check{G})$, the secrecy rate in (36) can be re-expressed as

$$R_{sopt}(\mathbf{G}) = \log_2\left(\frac{\left(\check{G}+\hat{G}(1+\bar{c}\ln\varepsilon)\right)\left((\bar{e}+\bar{d})(\check{G}+\hat{G})+1\right)}{\left(\bar{e}(\check{G}+\hat{G})+1\right)\left(\check{G}+\hat{G}(1+(\bar{c}-\bar{a})\ln\varepsilon)\right)}\right) \tag{37}$$

The transmission region for this case is $\{\mathbf{G}|\check{G} > \beta\}$, where $\beta$ is

$$\frac{1}{2\bar{d}}\left(-(2\hat{G}\bar{d}+\hat{G}\bar{e}\ln\varepsilon+\hat{G}\bar{c}\bar{d}\ln\varepsilon)+\sqrt{\hat{G}\ln\varepsilon(\hat{G}\ln\varepsilon\bar{a}^2\bar{e}^2+2\hat{G}\ln\varepsilon\bar{a}\bar{c}\bar{d}\bar{e}-4\bar{a}\bar{d}+\hat{G}\ln\varepsilon\bar{c}^2\bar{d}^2)}\right) \tag{38}$$

Please refer to Appendix F for the proof.

Accordingly, the maximum secrecy throughput for MRT plan is derived as

$$\Upsilon = \int_0^\infty \int_\beta^\infty \frac{e^{-x}x^{N_D-C-1}}{(N_{D-C}-1)!} \frac{e^{-y}y^{N_C-1}}{(N_C-1)!} \log_2\left(\frac{(x+y(1+\bar{c}\ln\varepsilon))\left((\bar{e}+\bar{d})(x+y)+1\right)}{(\bar{e}(x+y)+1)(x+y(1+(\bar{c}-\bar{a})\ln\varepsilon))}\right) dxdy \tag{39}$$

According to [42, 4.337.5], the above integral becomes as

$$\Upsilon = \sum_{m=0}^{N_D-N_C-1} \frac{1}{(N_{D-C}-m-1)!} \int_0^\infty f_{\hat{G}}(y) e^{-\beta}\beta^{N_D-N_C-m-1}[F(q_1)+F(q_2)-F(q_3)-F(q_4)+q_5]\,dy \tag{40}$$

where $F(q) = \frac{1}{\ln 2}\sum_{\mu=0}^{m}\frac{1}{(m-\mu)!}\left(\frac{(-1)^{m-\mu-1}}{q^{m-\mu}}e^{\frac{1}{q}}\text{Ei}\left[-\frac{1}{q}\right]+\sum_{n=0}^{m-\mu}(n-1)!\left(-\frac{1}{q}\right)^{m-\mu-n}\right)$, $q_1 = \frac{1}{\beta+y(1+\bar{c}\ln\varepsilon)}$, $q_2 = \frac{\bar{e}+\bar{d}}{(\bar{e}+\bar{d})(\beta+y)+1}$, $q_3 = \frac{1}{\beta+y(1+(\bar{c}-\bar{a})\ln\varepsilon)}$, $q_4 = \frac{\bar{e}}{\bar{e}(\beta+y)+1}$ and $q_5 = \log_2\left[\frac{[\beta+y(1+\bar{c}\ln\varepsilon)][(\bar{e}+\bar{d})(\beta+y)+1]}{[\beta+y(1+(\bar{c}-\bar{a})\ln\varepsilon)][\bar{e}(\beta+y)+1]}\right]$. Please refer to Appendix G for the proof.

To obtain a deeper insight, we proceed to investigate the secrecy throughput in *High-SNR Regime*. To do so, we consider the ideal hardware SNR in (8) and the definition of $R_s(\tau)$ in (27). As such, the secrecy rate is derived as

$$R_s^\infty(\tau) = \log_2\left(\frac{1+\mathcal{K}}{1+\tau k^\infty(\tau)}\right) \tag{41}$$

where $\mathcal{K} = \Upsilon_{D,AN}^\infty \triangleq \frac{1}{k_{tot}^2}$. Since $R_s^\infty(\tau) > 0$, the upper bound of $k^\infty(\tau)$ is acquired as $k_{max}^\infty = \frac{\mathcal{K}}{\tau}$. Using (9) and (12) we can conclude that

$$F_{Y_{E,AN}^{\infty}}(x) \triangleq \mathbb{P}_r\left\{\frac{\tau au}{(1-\tau)bv + \tau cu} < x\right\} = \mathbb{P}_r\left\{u < \frac{(1-\tau)bvx}{\tau(a-cx)}\right\}, \quad (42)$$

Using a similar process to (28), $Q^{\infty}(k^{\infty})$ is derived as

$$Q^{\infty}(k^{\infty}) = \left[1 + \frac{(1-\tau)k^{\infty}(\tau)}{a - c\tau k^{\infty}(\tau)}\right]^{-N_E-C} - \varepsilon \quad (43)$$

where $k^{\infty}(\tau)$ is a monotonically increasing function in the feasible set $0 \leq \tau \leq 1$, (Appendix H). Regarding (41) and its first derivative with respect to $\tau$ which is equal to $-\frac{1}{ln2} \times \frac{k^{\infty}(\tau) + \tau\frac{\partial k^{\infty}(\tau)}{\partial \tau}}{1+\tau k^{\infty}(\tau)} < 0$ and the monotonically increasing $k^{\infty}(\tau)$ with respect to $\tau$, we can conclude that $R_s^{\infty}(\tau)$ is monotonically decreasing with $\tau$. Thus, maximum secrecy rate at high-SNRs occurs for small $\tau$. This result states that most of the power $P$ should be allocated to AN signal while the remaining of the power is dedicated to the source for signal transmission. This is equivalent to the case where most of the power is allocated to AN in (18) where $\tau$ is very close to zero. Therefore, in high-SNR regime it is not easily possible for the eavesdropper to intercept the transmitted signal and therefore signal transmission is performed in most secure state.

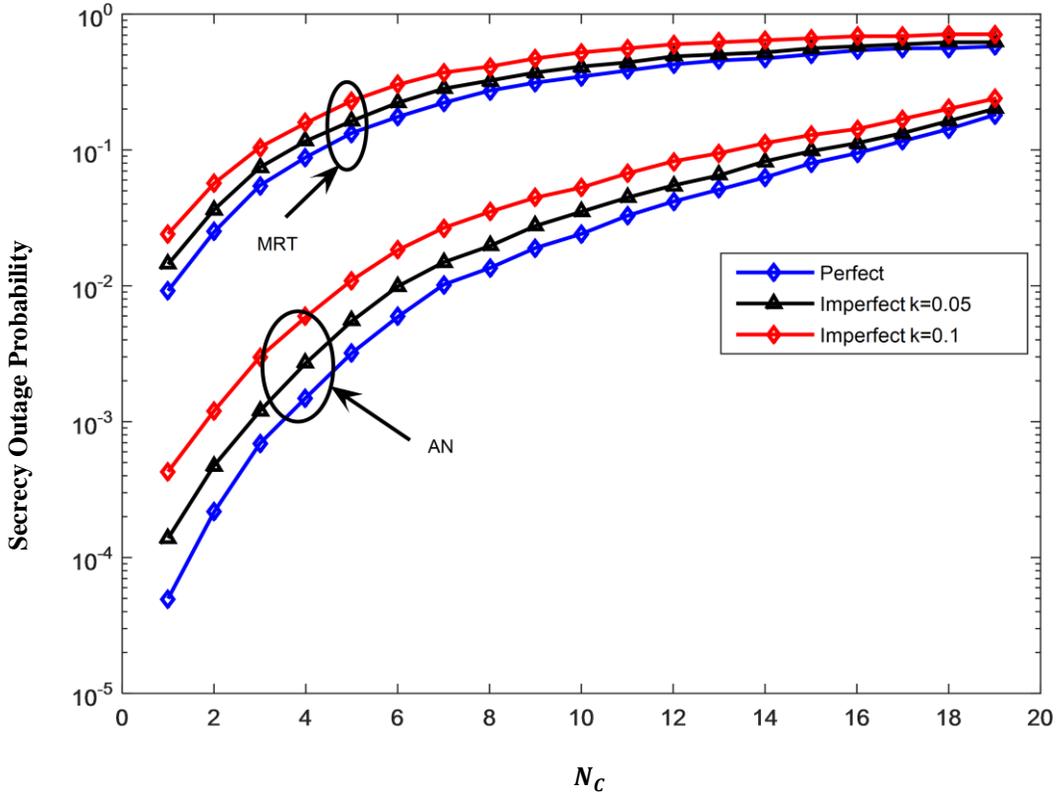

**Figure 3** SOP versus $N_C$ for different levels of hardware imperfections ($k_{tx} = k_{rx} = k$) with $M = 100$, $P = 5dBm$, $R_s = 5bits/s/Hz$, $N_D = 20$ and $d_D = 100m$

## 6. NUMERICAL EXAMPLES

In this section, numerical illustrations are presented to verify the accuracy of the analytical results. The transmitter is considered to be a linear array with antenna spacing of half wavelength. The path loss is modelled as $\alpha(dB) = a + b\,10\log_{10}(d)$ according to experimental results in [21] and the noise power is set to $\sigma_n^2 = -50$ dBm. In this model, $d$ is the distance in meters, $a$ and $b$ were measured at the carrier frequency of 28 GHz as 61.4 and 2 respectively. For comparison of the systems with perfect ($k = 0$) and imperfect hardware and also to investigate the imperfection level effect on the secrecy performance, we have considered two values of EVM as $k = 0.05$ and $k = 0.1$ which lies within the range mentioned in [27, Sec. 14.3.4] and also implemented in [28]. Transmitter is considered to be a ULA with antenna spacing half wavelength and the destination is located at $d_D = 100m$ from the source transmitter.

Fig. 3 compares the SOP of MRT and AN beamforming for different hardware imperfection levels and different numbers of common paths $N_C$. An eavesdropper is located at a distance $d_E = 100m$ from the transmitter and $M$ is set to 100. We can draw three conclusions based on what is observed in this figure. First, SOP is increased when $N_C$ increases. The reason is that increasing the number of common paths, the eavesdropper will have more opportunities to intercept information from the received signal to the destination. Second, increasing the imperfection level brings about a higher SOP. This is because the imperfection has a higher impact on the received SNDR at the destination compared to the eavesdropper which results in a higher eavesdropping probability. Third, AN beamforming significantly outperforms MRT because AN provides effective interference at the eavesdropper which makes it more advantageous for security.

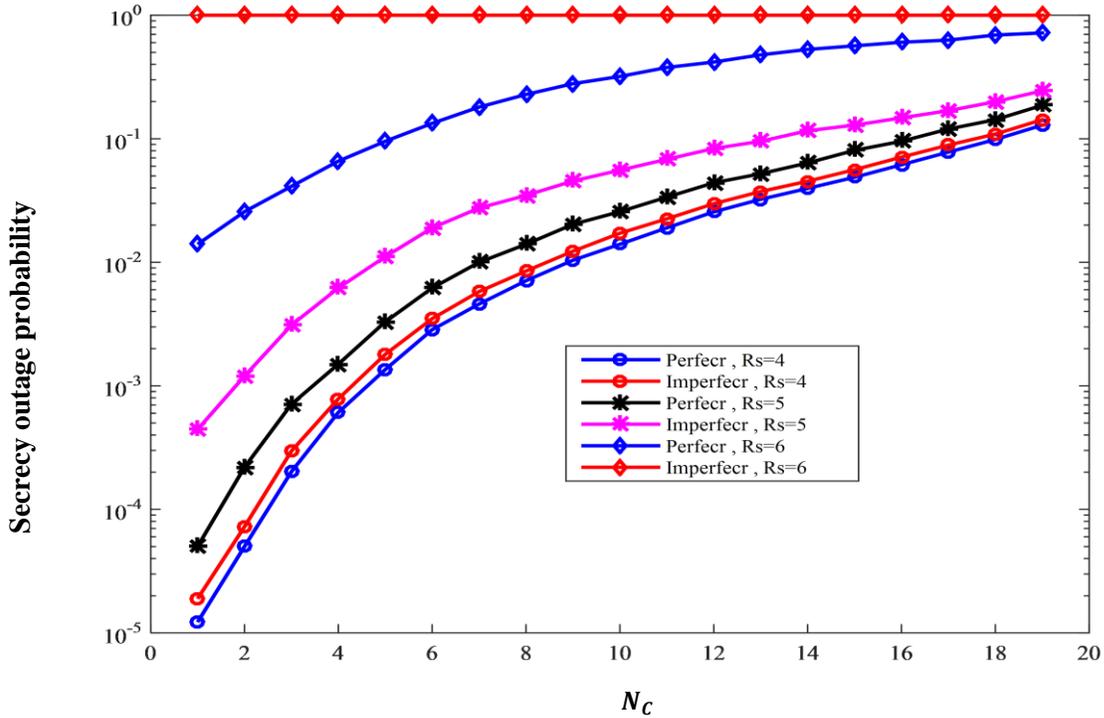

**Figure 4** SOP versus $N_C$ for different secrecy rates and hardware imperfection levels $k_{tx} = k_{rx} = 0.1$ with $P = 5dBm$, $N_D = 20$ and $d_D = 100m$

Fig. 4 shows the SOP versus $N_C$ for different amounts of secrecy rate. It is shown that as $R_s$ increases, the difference between SOP of the perfect and imperfect hardware becomes larger. It is further noted that for $R_s = 6 bits/s/Hz$ the system with imperfect hardware is always in secrecy outage which causes the communication to be insecure. This agrees with our analytical results in (20) which explains that the system is in complete secrecy outage for the derived threshold.

In Fig. 5, we plot the optimal power allocation $\tau^*_{opt}$ for different hardware imperfection levels and transmit powers $P$. The plot shows that $\tau^*_{opt}$ decreases with higher imperfection levels resulting in more AN transmitted at the source. We also see that $\tau^*_{opt}$ decreases with increasing $P$, which means that more AN signals should be transmitted when the information signal is successfully decoded at the destination.

Fig. 6 plots the secrecy throughput in terms of transmission power $P$. It is reasonable that increasing the transmission power results in higher secrecy throughput. However, as previously mentioned, hardware imperfection reduces the SNDR at the destination more than the eavesdropper which results in a lower secrecy throughput compared with the ideal hardware case. For further comparison, we have included the baseline performance of equal power allocation between the useful signal and artificial noise. The figure clearly shows that the optimal power allocation can achieve significantly higher secrecy throughput especially in the imperfect hardware scenario.

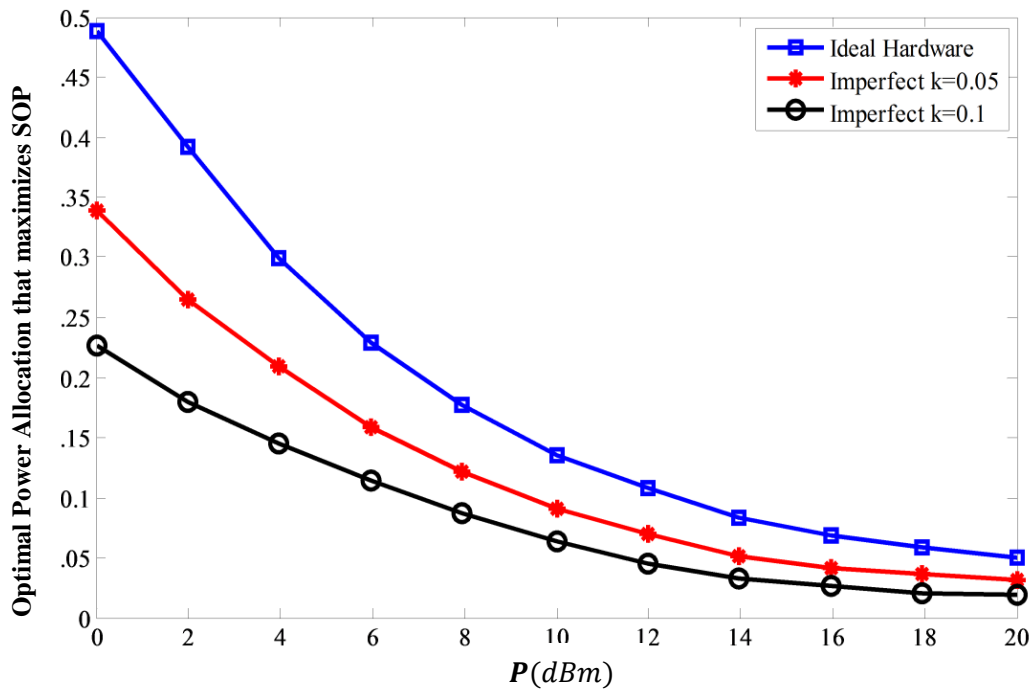

**Figure 5** OPA versus $P$ for different levels of hardware imperfections ($k_{tx} = k_{rx} = k$) with $M = 150$, $d_D = 100m$, $R_s = 5 bits/s/Hz$ and $N_D = 20$

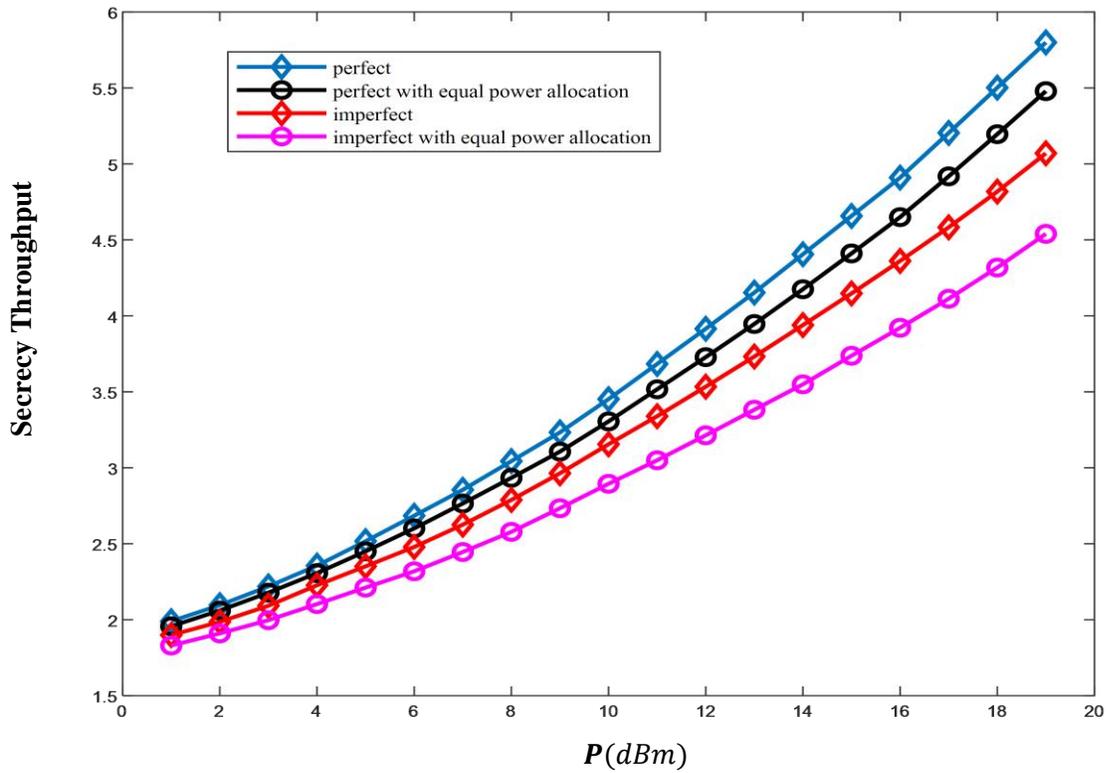

**Figure 6** Secrecy throughput in terms of $P$ for perfect and imperfect hardware, with $M = 100$, $d_D = 100m$, $N_D = 20$, $N_C = 16$ and $\varepsilon = 0.01$.

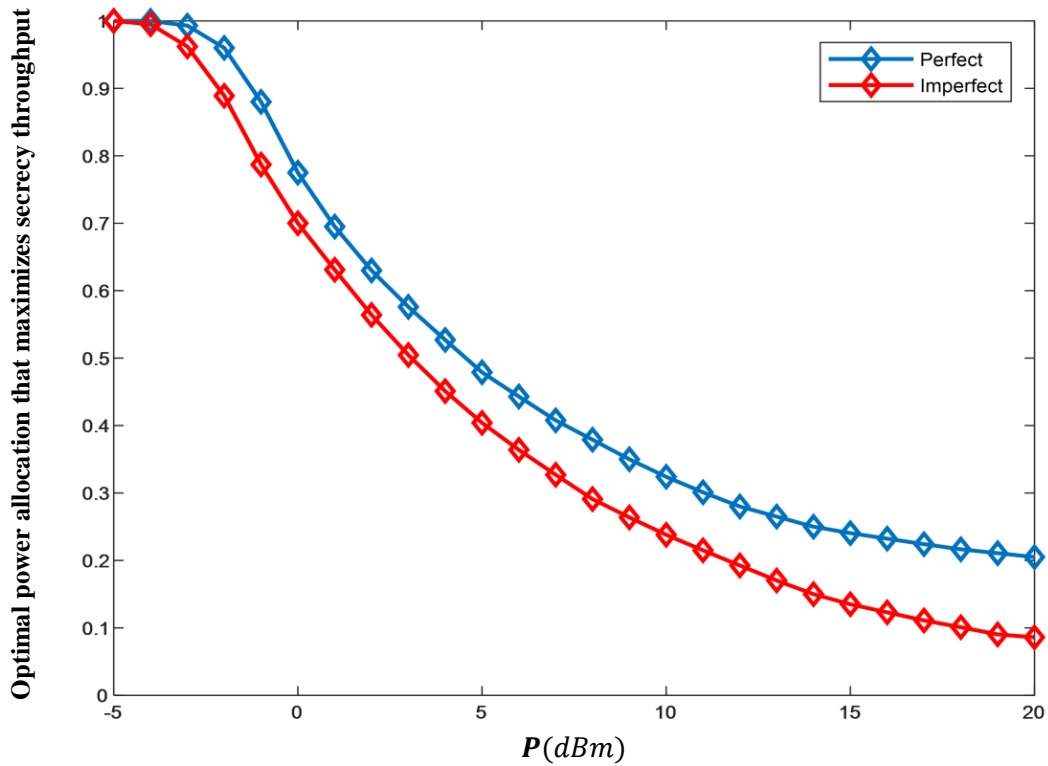

**Figure 7** Optimal power allocation factor that maximizes secrecy throughput in terms of $P$ for perfect and imperfect hardware with $M = 100$, $d_D = 100m$, $N_D = 20$ and $\varepsilon = 0.01$

Fig. 7 plots the optimal power allocation factor $\tau_{to}^*$ that maximizes the secrecy throughput for perfect and imperfect hardware cases. When the transmission power $P$ is very low, the power allocation factor is one because all the power should be dedicated to the information signal. By increasing $P$, the transmission link is guaranteed and therefore a reduction in $\tau_{to}^*$ means that the allocated power to AN increases to interfere with the eavesdropper. By comparing these two cases, we see that increasing hardware imperfection leads to a decrease in $\tau_{to}^*$ compared with ideal case. Thus, more power should be dedicated to AN to interfere with the eavesdropper and hence improve the secrecy performance of the system with imperfect hardware. In addition, when $P$ is high enough, we can observe that $\tau_{to}^*$ approaches zero as discussed in Section 5 for the high SNR regime.

## 7. CONCLUSION

In high rate mm-wave communications systems, physical transceiver hardware imperfections can create a remarkable distortion in the emitted and received signals. This paper examined the presence of hardware imperfections in mm-wave systems with multiple transmit antennas and slow-fading channels. The SOP and secrecy throughput is derived assuming on-off transmission and no knowledge of the eavesdropper CSI. Our results concluded that hardware imperfections and the number of common propagation paths can substantially decrease the secrecy performance. Optimal power allocation (OPA) solutions are derived to minimize the SOP and maximize the secrecy throughput. Simulation results highlight that the optimal power allocation factor varies according to the level of hardware imperfections which confirm the validity of our derived analytical results.

## APPENDIX A:

Three conditions have to be met for $\phi(\tau)$ to be considered as a concave function in the feasible set $(\tau_{min}, 1]$ (*corollary 1 in* [15]): i) $\left.\frac{\partial \phi(\tau)}{\partial \tau}\right|_{\tau=\tau_{min}} > 0$, ii) $\left.\frac{\partial \phi(\tau)}{\partial \tau}\right|_{\tau=1} < 0$, iii) There is only one maximum in the interval $(\tau_{min}, 1]$.

On the other hand, as discussed before, the positivity/negativity of $\frac{\partial \phi(\tau)}{\partial \tau}$ depends only on the numerator $\Omega(\tau)$. Along this line, by plugging the boundaries $\tau = \tau_{min}$ and $\tau = 1$ into $\Omega(\tau)$ we can find that the sign of both $\Omega(1)$ and $\Omega(\tau_{min})$ is unknown in the range of possible values of coefficients $\varepsilon_i|_{i=1}^3$. Thus, the following conclusions can be drawn for $\phi(\tau)$

i. If $\Omega(1).\Omega(\tau_{min}) > 0$, it is expected for $\phi(\tau)$ to be neither convex nor concave as can be seen in Fig.8 (A) and Fig.8 (B). Moreover, both roots of $\Omega(\tau)$ lie within the range of feasible set. So, in this situation the maximum of $\phi(\tau)$ may happen at $\tau = \tau_{min}$ or $\tau = 1$ or possibly at $\tau_1^\circ, \tau_2^\circ \in (0,1)$ which denote the roots of $\Omega(\tau)$ in the range of feasible set. The OPA factor $\tau_{opt}^*$ is selected by comparing $\Omega(\tau_{min}), \Omega(\tau_1^\circ), \Omega(\tau_2^\circ)$ and $\Omega(1)$.

ii. If $\Omega(\tau_{min}) < 0$, $\Omega(1) > 0$ we expect for $\phi(\tau)$ to be a convex function as can be illustrated in Fig.8 (C). Here, the maximum of $\phi(\tau)$ occurs at $\tau = \tau_{min}$ or $\tau = 1$ and the OPA factor $\tau_{opt}^*$ is chosen by comparing $\Omega(\tau_{min})$ and $\Omega(1)$.

iii. Finally if $\Omega(\tau_{min}) > 0$ & $\Omega(1) < 0$. In this situation, as shown in Fig. 8 (D), given that only $\tau_{opt}^* = \tau_1^\circ$ lies within the feasible set and the other root is outside of this interval, the corrolary 1 in [15] holds true and it can be found that $\phi(\tau)$ is a concave function in the feasible set.

## APPENDIX B:

From the definition of $\mathcal{P}_{SO}$ we have

$$\mathcal{P}_{SO} \triangleq \mathbb{P}_r\{C_E > R_t - R_s\} = \mathbb{P}_r\{Y_{E,AN} > 2^{R_t - R_s} - 1\} = 1 - F_{Y_{E,AN}}(2^{R_t - R_s} - 1), \tag{B1}$$

Using the constraint $\mathcal{P}_{SO} \leq \varepsilon$ and defining $k(\tau) \triangleq \frac{F_{Y_{E,AN}}^{-1}(1-\varepsilon)}{\tau}$, we have

$$R_s \leq R_t - \log_2(\tau k(\tau) + 1), \tag{B2}$$

Furthermore, since we have $R_t \leq C_D$, the upper bound of $R_s$ is acquired as

$$R_s(\tau) \triangleq C_D - \log_2(\tau k(\tau) + 1) = \log_2\left(\frac{1 + Y_{D,AN}}{\tau k(\tau) + 1}\right), \tag{B3}$$

Having $Y_{D,AN}$ in (6), we can reformulate (B3) given by

$$R_s(\mathbf{G}) \triangleq \log_2\left(\frac{\tau(d+e)+1}{\tau^2 ek(\tau) + \tau(k(\tau)+e)+1}\right), \tag{B4}$$

On the other hand to achieve positive secrecy rate, i.e., $R_s(\mathbf{G}) \geq 0$, the node $S$ transmits signal when $\mathbf{G}$ satisfies the transmission region's constraint $Y_{D,AN} \geq \tau k(\tau)$, resulting in transmission constraint as $\frac{d}{\tau e+1} \geq k(\tau)$.

**APPENDIX C:**

The first order derivative of $k(\tau)$ with respect to $\tau$ is calculated as below using the derivation criterion for implicit functions

$$\frac{dk}{d\tau} = -\frac{\partial Q/\partial \tau}{\partial Q/\partial k} = -\frac{A_1(k)}{A_2(k)} = -\frac{k_{tx}^2 k^2(a - c\tau k + (1-\tau)k) + k N_{E-C}(k_{tx}^2 k - 1)(a - c\tau k)}{(a - c\tau k + (1-\tau)k) + N_{E-C}(a - c\tau k)(1-\tau)}, \tag{C1}$$

Note that, $N_{E-C}, \tau, k, k_{tx}^2$ are positive variables. Additionally, to investigate the positivity/negativity of the term $(a - c\tau k)$ appeared in both the numerator and denominator, we obtain its lower bound by inserting $k_{max} = \frac{d}{\tau e+1}$ therein as follows.

$$a - c\tau k > a - c\tau k_{max} \xRightarrow{k_{max} = \frac{d}{\tau e+1}} a - c\tau k > a \cdot \left(\frac{1 + k_{rx}^2 \tau d}{\tau k_{tot}^2 d + 1}\right) > 0, \tag{C2}$$

Thus the denominator is always positive, i.e., $A_2(k) > 0$, and it can be easily found that the sign of $\frac{dk}{d\tau}$ follows that of the term $(k_{tx}^2 k - 1)$ in the numerator $A_1(k)$. On the other hand, for practical values of $\alpha_D, \alpha_E, \sigma_n^2$, and $k_{tx}$ we always have $\frac{1}{k_{tx}^2} \gg k_{max}$ in the feasible set $\tau \in [0,1]$. Therefore, the case $k > \frac{1}{k_{tx}^2}$ cannot occur in practice and the term $(k_{tx}^2 k - 1)$ will be always negative. Based on abovementioned discussion, it is easy to examine that $A_1(k)$ for practical values of different parameters therein is always negative, being responsible for the fact that $k$ is monotonically increasing function of $\tau$, i.e., $\frac{dk}{d\tau} > 0$.

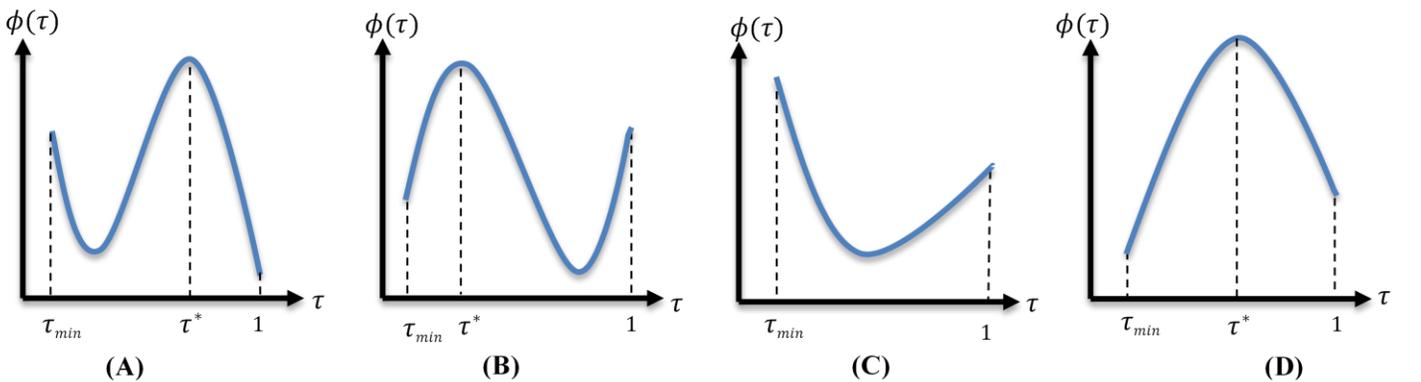

**Figure 8** Graphical representations of all possible cases of $\phi(\tau)$ versus $\tau$. Case I is illustrated in (A) and (B). Case II and Case III have been respectively, illustrated in (C) and (D).

## APPENDIX D:

Since $k$ is monotonically increasing by $\tau$, $k_{max}$ is obtained for $\tau = 1$. In addition, from (28) we know that, $Q(k(\tau = 1)) = 0$. Thus,

$$Q(k(\tau = 1)) = e^{-\frac{k(1)}{a-ck(1)}}[1] - \varepsilon = 0 \Rightarrow k_{max} = \frac{-a\ln\varepsilon}{1 + c\ln\varepsilon} \quad (D1)$$

But, $\frac{\partial Q}{\partial k}$ is

$$\frac{\partial Q}{\partial k} = \frac{-a}{(a - c\tau k)^2} e^{-\frac{k}{a-c\tau k}} \left[1 + \frac{(1-\tau)k}{a - c\tau k}\right]^{-(N_{E-C})}$$

$$- (N_{E-C}) e^{-\frac{k}{a-c\tau k}} \left[1 + \frac{(1-\tau)k}{a - c\tau k}\right]^{-(N_{E-C})-1} (1-\tau)a \quad (D2)$$

From APPENDIX C, we know that $(a - c\tau k) > 0$. Accordingly, using (D2) it is easily concluded that $\frac{\partial Q}{\partial k} < 0$.

## APPENDIX E:

Based on (27) and Proposition 1, we have $R_{S_{max}} = \log_2\left(\frac{1+Y_D}{\tau k(\tau)+1}\right)$. Thus, $\frac{dR_S}{d\tau} = \frac{1}{\ln 2}\left(\frac{1}{(\tau e+1)(\tau(e+d)+1)} - \frac{k+\tau\frac{dk}{d\tau}}{1+\tau k}\right)$. But, $\frac{d^2k}{d\tau^2}$ and in turn $\frac{d^2R_S}{d\tau^2}$ are very convoluted terms and the convexity of $k$ and $R_S$ with respect to $\tau$ is not straightforward to analyze mathematically. Thus, we proceed to consider the following cases. First, $\frac{dR_S}{d\tau} = \frac{1}{\ln 2}(d - k)$ for $\tau = 0$. Since $\frac{d}{\tau e+1} \geq k(\tau)$, then $d > k$ and hence $\frac{dR_S}{d\tau} > 0$, $\tau = 0$. Therefore, $R_S$ is a nonconvex function of $\tau$ in the interval $0 < \tau < 1$. If $\frac{d^2R_S}{d\tau^2} > 0$, then $R_S$ is a concave function of the $\tau$. When $\tau = 1$, which is equivalent with $k = k_{max} = -a\ln(\varepsilon)/(1 + c\ln(\varepsilon))$, $\frac{dR_S}{d\tau}$ is

$$\frac{dR_S}{d\tau} = \frac{d}{(e+1)(e+d+1)} + \frac{(N_{E-C} + 1)a\ln\varepsilon\left(1 + \frac{c\ln\varepsilon}{1+c\ln\varepsilon}\right)}{1 + (c-a)\ln\varepsilon} \quad , \tau = 1 \quad (E3)$$

In this case in which $R_S$ is a concave function of $\tau$, if $\frac{dR_S}{d\tau} > 0$ for $\tau = 1$, then maximum of $R_S$ will occur at $\tau = 1$. If $\frac{dR_S}{d\tau} < 0$ for $\tau = 1$, then $R_S$ will have an extremum at the root of the equation $\frac{dR_S}{d\tau} = 0$. When $R_S$ is non-concave, depending on sign of $\frac{dR_S}{d\tau}$ at $\tau = 1$ and based on Fig. 9, two possible cases are $\frac{dR_S}{d\tau} > 0$ and $\frac{dR_S}{d\tau} < 0$ for $\tau = 1$. Therefore, maximum amount of $R_s(\tau)$ is

$$\max R_s(\tau) = \begin{cases} \max(R_s(\tau_1), R_s(1)), & Z > 0 \\ \max(R_s(\tau_1'), R_s(\tau_3)), & otherwise \end{cases} \quad (E4)$$

Hence, $\tau_{opt}$ could be derived as (33).

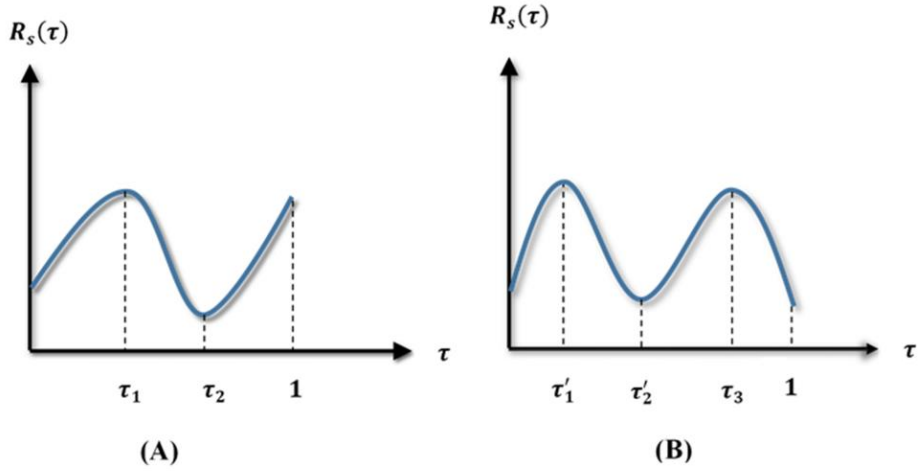

**Figure 9** Graphical representations of all possible cases of $R_s(\tau)$ versus $\tau$ when $R_s(\tau)$ is non-concave. (A) represents case 1 in which $\frac{dR_s}{d\tau} > 0$ for $\tau = 1$. (B) case 2 in which $\frac{dR_s}{d\tau} < 0$ for $\tau = 1$.

## APPENDIX F:

$$\xi_{to} = \left\{ \mathbf{G} \bigg| \frac{d}{\tau e + 1} > k \right\} = \left\{ \mathbf{G} \bigg| \frac{\bar{d}(\hat{G} + \check{G})}{\bar{e}(\hat{G} + \check{G}) + 1} > k \right\} \quad (F1)$$

For MRT plan, $\tau = 1$ and $k = k_{max} = -a\ln(\varepsilon)/(1 + c\ln(\varepsilon))$. Therefore, $\frac{\bar{d}(\hat{G}+\check{G})}{\bar{e}(\hat{G}+\check{G})+1} > k$ becomes as

$$\frac{\bar{d}(\hat{G} + \check{G})}{\bar{e}(\hat{G} + \check{G}) + 1} > \frac{-\bar{a}\frac{\hat{G}\ln\varepsilon}{\hat{G}+\check{G}}}{1 + \bar{c}\frac{\hat{G}\ln\varepsilon}{\hat{G}+\check{G}}} = \frac{-\bar{a}\hat{G}\ln\varepsilon}{\hat{G} + \check{G} + \bar{c}\hat{G}\ln\varepsilon} \quad (F2)$$

From $(F2)$ we can conclude that

$$\check{G}^2 + \check{G}\left(\hat{G}\left(2 + \bar{c}\ln\varepsilon + \frac{\bar{a}\bar{e}}{\bar{d}}\ln\varepsilon\right)\right) + \hat{G}^2\left(1 + \bar{c}\ln\varepsilon + \frac{\bar{a}\bar{e}}{\bar{d}}\ln\varepsilon\right) + \frac{\bar{a}}{\bar{d}}\hat{G}\ln\varepsilon > 0 \quad (F3)$$

$(F3)$ is similar to $x^2 + a_1 x + a_2 > 0$ where $x$ is substituted by $\check{G}$, $a_1$ by $\hat{G}\left(2 + \bar{c}\ln\varepsilon + \frac{\bar{a}\bar{e}}{\bar{d}}\ln\varepsilon\right)$ and $a_2$ by $\hat{G}^2\left(1 + \bar{c}\ln\varepsilon + \frac{\bar{a}\bar{e}}{\bar{d}}\ln\varepsilon\right) + \frac{\bar{a}}{\bar{d}}\hat{G}\ln\varepsilon$. Since $\check{G} \triangleq \|\check{\mathbf{g}}_D\|^2$ is positive, deriving the roots of the equation $x^2 + a_1 x + a_2$ for substituted $x$, $a_1$ and $a_2$, we can conclude that $(F3)$ holds for the case of $(38)$.

## APPENDIX G:

Defining variable $\bar{x} = x - \beta$, (39) becomes as

$$\int_0^\infty \int_0^\infty \frac{e^{-(\bar{x}+\beta)}(\bar{x}+\beta)^{N_{D-C}-1}}{(N_{D-C}-1)!} f_{\hat{G}}(y) \log_2 \frac{\left((\bar{x}+\beta) + y(1+\bar{c}\ln\varepsilon)\right)\left((\bar{e}+\bar{d})((\bar{x}+\beta) + y) + 1\right)}{\left(\bar{e}((\bar{x}+\beta) + y) + 1\right)((\bar{x}+\beta) + y(1 + (\bar{c}-\bar{a})\ln\varepsilon))} dxdy \quad (G1)$$

we can replace $\log_2 \frac{((\bar{x}+\beta)+y(1+\bar{c}\,ln\varepsilon))((\bar{e}+\bar{d})((\bar{x}+\beta)+y)+1)}{(\bar{e}((\bar{x}+\beta)+y)+1)((\bar{x}+\beta)+y(1+(\bar{c}-\bar{a})ln\varepsilon))}$ by $\frac{1}{ln2}[ln(R_1) + ln(R_2) - ln(R_3) - ln(R_4) + ln(R_5)]$ in which $R_1 = 1 + \frac{1}{\beta+y(1+\bar{c}\,ln\varepsilon)}\bar{x}$, $R_2 = 1 + \frac{(\bar{e}+\bar{d})}{(\bar{e}+\bar{d})(\beta+y)+1}\bar{x}$, $R_3 = 1 + \frac{1}{\beta+y(1+(\bar{c}-\bar{a})ln\varepsilon)}\bar{x}$, $R_4 = 1 + \frac{\bar{e}}{\bar{e}(\beta+y)+1}\bar{x}$ and $R_5 = \frac{[\beta+y(1+\bar{c}\,ln\varepsilon)][(\bar{e}+\bar{d})(\beta+y)+1]}{[\beta+y(1+(\bar{c}-\bar{a})ln\varepsilon)][\bar{e}(\beta+y)+1]}$. Also $(\bar{x}+\beta)^{N_D-C-1} = \sum_{m=0}^{N_D-C-1}\binom{N_D-C-1}{m}\beta^{N_D-C-m-1}\bar{x}^m$.

Thus, (G1) is rewritten as

$$Y = \frac{1}{ln2}\int_0^\infty \frac{e^{-\beta}f_{\hat{G}}(y)}{(N_D-N_C-1)!}\sum_{m=0}^{N_D-C-1}\binom{N_D-C-1}{m}\beta^{N_D-C-m-1}\int_0^\infty \left[ln\left(1+\frac{1}{\beta+y(1+\bar{c}\,ln\varepsilon)}\bar{x}\right)\right.$$

$$+ ln\left(1+\frac{(\bar{e}+\bar{d})}{(\bar{e}+\bar{d})(\beta+y)+1}\bar{x}\right) - ln\left(1+\frac{1}{\beta+y(1+(\bar{c}-\bar{a})ln\varepsilon)}\bar{x}\right)$$

$$\left. - ln\left(1+\frac{\bar{e}}{\bar{e}(\beta+y)+1}\bar{x}\right) + ln\left(\frac{[\beta+y(1+\bar{c}\,ln\varepsilon)][(\bar{e}+\bar{d})(\beta+y)+1]}{[\beta+y(1+(\bar{c}-\bar{a})ln\varepsilon)][\bar{e}(\beta+y)+1]}\right)\right]\bar{x}^m e^{-\bar{x}}d\bar{x}dy$$

(G2)

$\frac{1}{(N_D-C-1)!}\binom{N_D-C-1}{m}$ is replaced by $\frac{1}{m!(N_D-C-m-1)!}$. Therefore, (G2) becomes as

$$\sum_{m=0}^{N_D-N_C-1}\frac{1}{m!(N_D-C-m-1)!}\int_0^\infty e^{-\beta}f_{\hat{G}}(y)\beta^{N_D-C-m-1}$$

$$\times \left(\int_0^\infty [ln(1+q_1\bar{x}) + ln(1+q_2\bar{x}) - ln(1+q_3\bar{x}) - ln(1+q_4\bar{x}) + q_5]\bar{x}^m e^{-\bar{x}}d\bar{x}\right)$$

(G3)

Using [42, 4.337.5], (G3) is changed to (40).

**APPENDIX H:**

$$\frac{dk^\infty}{d\tau} = -\frac{\partial Q^\infty/\partial \tau}{\partial Q^\infty/\partial k^\infty} = -\frac{kbc^{\infty 2}\tau^2 - abk^\infty \tau^2}{ab\tau^2 - ab\tau^3} = \frac{abk^\infty - bck^{\infty 2}}{ab - ab\tau} \tag{H1}$$

Since $0 < \tau < 1$, denominator is positive. The numerator is equal to $abk^\infty - bck^{\infty 2} = abk^\infty(1-k^\infty k_{tx}^2)$ because $c = k_{tx}^2 a$. Since $a$, $b$ and $k^\infty$ are positive, $k^\infty$ is monotonically increasing by $\tau$ when $k^\infty < \frac{1}{k_{tx}^2} < k_{max}^\infty = \frac{\mathcal{K}}{\tau}$.